\documentclass[twocolumn,10pt,a4]{IEEEtran}%
\usepackage[utf8]{inputenc}
\usepackage{amsfonts}
\usepackage{amsmath}
\usepackage{multicol}
\usepackage{amssymb}
\usepackage{graphicx}
\usepackage{subcaption}
\usepackage{xcolor}
\usepackage{cite}

\usepackage{cite}%
\providecommand{\U}[1]{\protect\rule{.1in}{.1in}}
\providecommand{\U}[1]{\protect\rule{.1in}{.1in}}
\providecommand{\U}[1]{\protect\rule{.1in}{.1in}}
\providecommand{\U}[1]{\protect\rule{.1in}{.1in}}
\providecommand{\U}[1]{\protect\rule{.1in}{.1in}}
\providecommand{\U}[1]{\protect\rule{.1in}{.1in}}
\providecommand{\U}[1]{\protect\rule{.1in}{.1in}}
\providecommand{\U}[1]{\protect\rule{.1in}{.1in}}
\providecommand{\U}[1]{\protect\rule{.1in}{.1in}}
\providecommand{\U}[1]{\protect\rule{.1in}{.1in}}

\begin{document}

\title{The Support Uncertainty Principle and the Graph Rihaczek Distribution: 
	Revisited and Improved}
	\author{
		Ljubi\v{s}a~Stankovi\'{c},~\IEEEmembership{Fellow,~IEEE}
		\thanks{
			L. Stankovi\'{c} is with the  University of
			Montenegro, 81000 Podgorica, Montenegro.
			Contact e-mail: ljubisa@ac.me.
			% The author is thankful to Prof. Milo\v{s} Dakovi\'{c} and Dr. Milo\v{s} Brajovi\'{c} for constructive comments on the manuscript.
		}
	}
	\maketitle

	\begin{abstract}

The classical support uncertainty principle states that the signal and its discrete Fourier transform (DFT) cannot be localized simultaneously in an arbitrary small area in the time and the frequency domain. The product of the number of nonzero samples in the time domain and the frequency domain is greater or equal to the total number of signal samples. The support uncertainty principle has been extended to the arbitrary orthogonal pairs of signal basis and the graph signals, stating that the product of supports in the vertex domain and the spectral domain is greater than the reciprocal squared maximum absolute value of the basis functions. This form is then used in compressive sensing and sparse signal processing to define the reconstruction conditions. In this paper, we will revisit the graph signal uncertainty principle using the graph Rihaczek distribution as an analysis tool and derive an improved bound for the support uncertainty principle of graph signals.  
		\end{abstract}

\section{Introduction}

    The uncertainty principle is one of the signal processing keystones. The basic form of the uncertainty principle was originally established in  quantum mechanics  and is called the Robertson-Schr{\"o}dinger inequality. This form was used in classical time-frequency analysis to establish the lower bound for the product of effective signal widths (variances) in the time and the frequency domain \cite{boashash2015time,cohen1995time,stankovic1997highly}, and to show that an ideal localization in both time and frequency is not possible. Another form of this principle is the support uncertainty principle,  defined as a bound for the product of the signal and its transform supports. This form of the uncertainty principle is closely related to the sparsity support measures \cite{donoho2006compressed,ricaud2014survey, stankovic2001measure}, and it is commonly referred to as the support uncertainty principle. It plays a fundamental role not only in time-frequency analysis but also in compressive sensing and sparse signal processing.  Surveys of various forms of the uncertainty principle in signal analysis can be found in \cite{ricaud2014survey,perraudin2018global}.

Classical Fourier-based analysis support uncertainty principle was extended to the pairs of bases, directly applicable to graph signal processing  and compressive sensing, in \cite{elad2001generalized}. The uncertainty principle was considered and used in various graph signal processing approaches, including the joint vertex-frequency domain analysis, in \cite{pasdeloup2019uncertainty,Tsitsvero2016,Agaskar}.

In this paper, we shall revisit the graph signal uncertainty principle using the graph Rihaczek distribution \cite{stankovic2019vertexTEL} as an analysis tool and derive an improved bound for the support uncertainty principle of graph signals. The theory is illustrated on examples.

\section{Basic Definitions}

A graph is defined by $N$ vertices, denoted here by $n \in\mathcal{V}=\{0,1,\dots, N-1\}$. The vertices are connected with edges whose weights are $W_{mn}$.  For the vertices $m$ and $n$ that are not connected, $W_{mn}=0$ holds. The edge weights $W_{mn}$ are written in a matrix form, using the weight matrix $\mathbf{W}$. The graph is unweighted if all nonzero elements in the weight matrix are equal to unity. This specific form of the weight matrix is called the adjacency matrix and denoted by $\mathbf{A}$. The graph Laplacian is defined by $\mathbf{L}=\mathbf{D}-\mathbf{W}$, where $\mathbf{D}$ is a diagonal degree matrix $\mathbf{D}$, whose elements $D_{nn}$ are equal to the sum of all edge weights connected to the considered vertex, $n$. The Laplacian of an undirected graph is symmetric, $\mathbf{L}=\mathbf{L}^T$.  

Spectral analysis of graphs is most commonly based on the eigendecomposition of the graph Laplacian, $\mathbf{L}$, or the adjacency matrix, $\mathbf{A}$ \cite{stankovic2019graph}. 
By default, we will assume the decomposition of the graph Laplacian, if not stated otherwise. The eigenvectors, $\mathbf{u}_k$, and the eigenvalues, $\lambda_k$, of the graph Laplacian are calculated based on the usual definition 
\begin{gather}\mathbf{L}\mathbf{u}_k=\lambda_k \mathbf{u}_k,\end{gather} for $k=0,1,\dots,N-1$. Matrix form of this equation is 
\begin{gather}\mathbf{U}^{-1}\mathbf{L}\mathbf{U}=\boldsymbol{\Lambda},\end{gather}
where  $\mathbf{U}$ is the transformation matrix with eigenvectors  $\mathbf{u}_k$, $k=0,1,\dots,N-1$, as its columns, $u_k(n)$ being its elements, and $\boldsymbol{\Lambda}$ is a diagonal matrix with the elements $\lambda_k$.

A graph signal $x(n)$, $n=0,1,\dots,N-1$, is a set of data $x(n)$ associated with the vertices, as the signal domain.  

The graph Fourier transform (GFT) of a signal $\mathbf{x}=[x(0), \ x(1),\dots, x(N-1)]^T$ is defined by
\begin{gather}\mathbf{X}=\mathbf{U}^{-1}\mathbf{x},\end{gather}
where $\mathbf{X}=[X(0), \ X(1),\dots, X(N-1)]^T$ is the GFT vector with elements $X(k)$. The inverse GFT is defined by 
\begin{gather}\mathbf{x}=\mathbf{U}\mathbf{X}.\end{gather}

A  special case of a graph is the circular \textit{directed} and unweighted graph, when the sampling instants $n=0,1,\dots,N-1$ play the role of vertices. For this graph and the adjacency matrix, $\mathbf{A}$, the eigendecomposition results in the standard DFT basis functions (eigenvectors) 
\begin{gather}u_k(n)=\frac{1}{\sqrt{N}}e^{j2 \pi n k/N}, \ \ n=0,1,\dots,N-1, \end{gather} $k=0,1,\dots,N-1$, and classical Fourier analysis follows as a special case of the GFT analysis.  

\section{Graph Energy Distribution}

The graph Rihaczek distribution is defined by \cite{stankovic2018vertex}
\begin{gather}
E(n,k)=x(n)X(k)u_k(n).
\end{gather}
Without loss of generality, assume the unit signal energy,
\begin{gather}
E_x=\sum_{n=0}^{N-1}|x(n)|^2=\sum_{k=0}^{N-1}|X(k)|^2=1.
\end{gather}
The graph Rihaczek distribution satisfies the energy property,
\begin{gather}
\sum_{n=0}^{N-1}\sum_{k=0}^{N-1}E(n,k)=\sum_{n=0}^{N-1}\sum_{k=0}^{N-1}x(n)X(k)u_k(n) \nonumber \\
=\sum_{n=0}^{N-1}|x(n)|^2=\sum_{k=0}^{N-1}|X(k)|^2=E_x=1. \label{energy}
\end{gather}
This distribution satisfies the marginal properties as well \cite{stankovic2018vertex}.

% This distribution satisfies both the frequency marginal property,
%\begin{gather}
%\sum_{n=0}^{N-1}E(n,k)=\sum_{n=0}^{N-1}x(n)X(k)u_k(n)=X^2(k),
%\end{gather}
%and the vertex marginal property,
%\begin{gather}
%\sum_{k=0}^{N-1}E(n,k)=\sum_{k=0}^{N-1}x(n)X(k)u_k(n)=x^2(n).
%\end{gather}
%
 
The classical Rihaczek distribution is obtained within the DFT framework as
\begin{gather}
E(n,k)=x(n)X^*(k)u^*_k(n)\!=\!x(n)X^*(k)\tfrac{1}{\sqrt{N}}e^{-j2 \pi \tfrac{n k}{N}}. \label{ClassRD}
\end{gather}
It satisfies classical condition for the unbiased signal energy, and although it does not satisfy the nonnegativity property at each point $(n,k)$, it is called a distribution. 

\section{Support Uncertainty Principle Derivation}

From the Rihaczek distribution energy condition (\ref{energy}) follows 
\begin{gather}
1 \le \sum_{n=0}^{N-1}\sum_{k=0}^{N-1}|E(n,k)|=\sum_{n=0}^{N-1}\sum_{k=0}^{N-1}|x(n)| \ |X(k)| \ |u_k(n)|  \nonumber \\
 \le  \max_{n,k}\{|u_k(n)|\}\sum_{n=0}^{N-1}\sum_{k=0}^{N-1}|x(n)| \ |X(k)|.
\label{RDGun}
\end{gather}
This relation means that the $\ell_1$-norm of the Rihaczek distribution is lower or equal to the product of the $\ell_1$-norms of the signal and its GFT, $||\mathbf{x}||_1||\mathbf{X}||_1$,  multiplied by the maximum absolute element, $\max_{n,k}\{|u_k(n)|\}$, of the transformation matrix,  $\mathbf{U}$.

%, that is 
%\begin{gather}
%1 \le ||\mathbf{E}||_1\le \max_{n,k}\{|u_k(n)|\}||\mathbf{x}||_1||\mathbf{X}||_1, \label{ineq1}
%\end{gather}
%where $\mathbf{E}$ is the matrix with elements $E(n,k)$.
%Based on the well-known relation between the norm-one and norm-two ($||\mathbf{x}||_1 \le \sqrt{N}||\mathbf{x}||_2$) we can also write
%\begin{gather}
%||\mathbf{E}||_1\le \max_{n,k}\{|u_k(n)|\}||\mathbf{x}||_1||\mathbf{X}||_1
%\nonumber \\
%\le \max_{n,k}\{|u_k(n)|\} N||\mathbf{x}||_2||\mathbf{X}||_2=N \max_{n,k}\{|u_k(n)|\}.
%\end{gather}

%\medskip

%\noindent\textbf{Finite support signal and GFT.} 

Next, we will assume, as in \cite{elad2001generalized}, that the support $\mathbb{M}$ of the signal $x(n)$ is finite, 
\begin{gather}\mathbb{M}=\{n_1,n_2,\dots,n_M\},\end{gather} 
meaning that $x(n) \ne 0$ for $n \in \mathbb{M}$ and $x(n) = 0$ for $n \notin \mathbb{M}$, while the support of the graph Fourier transform $X(k)$ is 
\begin{gather}\mathbb{K}=\{k_1,k_2,\dots,k_K\},
\end{gather} 
where $X(k) \ne 0$ for $k \in \mathbb{K}$ and $X(k) = 0$ for $k \notin \mathbb{K}$. 
By definition, we can write the relations
\begin{gather}||\mathbf{x}||_0=\mathrm{card}\{\mathbb{M}\}=M \text{ \ and \ } ||\mathbf{X}||_0=\mathrm{card}\{\mathbb{K}\}=K.
\end{gather} 
Applying the Schwartz inequality to (\ref{RDGun}) squared, we get
\begin{gather}
1\!=\! \Big(\!\sum_{n \in \mathbb{M}}\sum_{k \in \mathbb{K}}E(n,k)\Big)^2 \!\! \le \!\! \Big(\sum_{n \in \mathbb{M}}\sum_{k \in \mathbb{K}}|x(n)| \ |X(k)| \ |u_k(n)|\Big)^2 \nonumber \\
= \Big(\sum_{n \in \mathbb{M}}\sum_{k \in \mathbb{K}} (\sqrt{|u_k(n)|}|x(n)|) \ (\sqrt{|u_k(n)} |X(k)|) \Big)^2  \label{lwbound} \\ \le \sum_{n \in \mathbb{M}}\sum_{k \in \mathbb{K}} |u_k(n)||x(n)|^2 \sum_{n \in \mathbb{M}}\sum_{k \in \mathbb{K}} |u_k(n)||X(k)|^2 \label{UnCElad0} \\
\le
 \max_{n,k}\{|u_k(n)|^2\} KM 
 = \max_{n,k}\{|u_k(n)|^2\} ||\mathbf{x}||_0 ||\mathbf{X}||_0, \label{UnCElad}
\end{gather}
since the unit energy of the graph signal is assumed, that is, $\sum_{n \in \mathbb{M}}|x(n)|^2=\sum_{k \in \mathbb{K}} |X(k)|^2=1$.

The  inequality in (\ref{UnCElad}) results in the support uncertainty principle \cite{elad2001generalized} 
\begin{gather}
||\mathbf{x}||_0 ||\mathbf{X}||_0 \ge \frac{1}{\displaystyle{\max_{n,k}}\{|u_k(n)|^2\}}. \label{uncMM}
\end{gather}
 
From the classical Rihaczek distribution (\ref{ClassRD}), with $\max_{n,k}\{|u_k(n)|^2\}=1/N$, the standard DFT support uncertainty principle follows
\begin{gather}
||\mathbf{x}||_0 ||\mathbf{X}||_0 \ge N. \label{ClassUP}
\end{gather}

\section{Improved Lower Bound}

The lower bound of the support uncertainty principle is calculated using the maximal absolute value of the basis functions, $\max_{n,k}\{|u_k(n)|\}$, for all $n \in \mathbb{M}$ and $k \in \mathbb{K}$. The support uncertainty principle bound can be improved by using a different grouping in the Schwartz inequality
%
%. The equality in this inequality holds if   
%$$\sqrt{|u_k(n)|}|x(n)|=\sqrt{|u_k(n)|}|X(k)|$$ 
%within the support for $n$ and $k$, $n \in \mathbb{M}$ and $k \in \mathbb{K}$. This is true if $|x(n)|=c|X(k)|$. Having in mind the energy condition for the signal,  we may use, as in \cite{elad2001generalized}, $|x(n)|=1/\sqrt{M}$ and $|X(k)|=1/\sqrt{K}$. This means that, starting from the first line in (\ref{UnCElad}), but using the Swartz relation equality condition, we can write
\begin{gather}
1\le \Big(\sum_{n \in \mathbb{M}}\sum_{k \in \mathbb{K}}|x(n)| \ |X(k)| \ |u_k(n)|\Big)^2 \nonumber \\
\le \! \sum_{n \in \mathbb{M}}\sum_{k \in \mathbb{K}}(|x(n)||X(k)|)^2 \! \sum_{n \in \mathbb{M}}\sum_{k \in \mathbb{K}}|u_k(n)|^2 \!=\!\sum_{n \in \mathbb{M}}\sum_{k \in \mathbb{K}} |u_k(n)|^2   \nonumber \\  =MK
\Big(\frac{1}{MK}\sum_{n \in \mathbb{M}}\sum_{k \in \mathbb{K}} |u_k(n)|^2\Big) 
=||\mathbf{x}||_0 ||\mathbf{X}||_0 \displaystyle{\mathrm{Avg}}\{|u_k(n)|^2\}
\label{ImUB}.
\end{gather}
This  means that, for any support sets $\mathbb{M}$ and $\mathbb{K}$, holds  
\begin{gather}
||\mathbf{x}||_0 ||\mathbf{X}||_0 \ge \frac{1}{\displaystyle{\mathrm{Avg}}\{|u_k(n)|^2\}}  \label{NewUP1} \\
=\frac{1}{\displaystyle{\frac{1}{||\mathbf{x}||_0 ||\mathbf{X}||_0}\sum_{n \in \mathbb{M}}\sum_{k \in \mathbb{K}} |u_k(n)|^2}}. \nonumber
\end{gather}
In general, the  inequality in (\ref{NewUP1}) is signal-dependent.
The sum of $|u_k(n)|$ in the last equation is always smaller or equal to the sum of $MK$ largest values of $|u_k(n)|$, denoted by $s(p)$. Therefore, we can write
\begin{gather}
||\mathbf{x}||_0 ||\mathbf{X}||_0 
\ge \frac{1}{\frac{1}{||\mathbf{x}||_0 ||\mathbf{X}||_0}\sum_{p=1}^{||\mathbf{x}||_0 ||\mathbf{X}||_0} s^2(p)}
, \label{NewUP2}
\end{gather}
where  
\begin{gather*}s(p)=\mathrm{sort}_{n,k}\{|u_k(n)|\},
\end{gather*}
with $n, k =0,1,\dots,N-1$, and $p=1,2,\dots, N^2$, 
are the values of  of $|u_k(n)|$ sorted into a \textit{nonincreasing order}.

\smallskip

\noindent\textbf{Illustrative example.} We shall present a simple direct search solution to (\ref{NewUP2}) using the sorted  values of $|u_k(n)|$  from Example 2 in Section \ref{SecNE}, 
$$\mathbf{s}=[0.5857, \ 0.5285, \   0.3743,  \  0.3669, \   0.3659, \ 0.3658, \  \dots ].$$

We start the calculation and check possible $||\mathbf{x}||_0 ||\mathbf{X}||_0=1$ bound. Replacing this value od $||\mathbf{x}||_0 ||\mathbf{X}||_0$ into (\ref{NewUP2}) we get 
$$1\ge \frac{1}{0.5857^2}=2.9156$$
which is obviously not true. Therefore, we cannot get the bound with the maximal value of $|u_k(n)|=s(1)=0.5857$. 
Then, we try with the next possible smallest bound with two elements, $||\mathbf{x}||_0 ||\mathbf{X}||_0=2$, in  (\ref{NewUP2}),  and get 
$$2\ge \frac{1}{\frac{1}{2}(0.5857^2+0.5285^2)}=3.2137.$$
Obviously, the bound cannot be obtained with $||\mathbf{x}||_0 ||\mathbf{X}||_0=2$. 
Next, we continue with $||\mathbf{x}||_0 ||\mathbf{X}||_0=3$, and $||\mathbf{x}||_0 ||\mathbf{X}||_0=4$ and we conclude that the corresponding inequalities do not hold. For  $||\mathbf{x}||_0 ||\mathbf{X}||_0=5$, we get 
$$5\ge \frac{1}{\frac{1}{5}(0.5857^2+0.5285^2+\dots+0.3659^2)}=4.8499.$$
This is the lowest  value of $||\mathbf{x}||_0 ||\mathbf{X}||_0$ producing the inequality which is true, meaning that the uncertainty principle is 
 $$||\mathbf{x}||_0 ||\mathbf{X}||_0 \ge 4.8499.$$

\noindent \textbf{Solution existence.} This iterative procedure always has a solution within $1 \le ||\mathbf{x}||_0 ||\mathbf{X}||_0 \le N$, since the expression  on the right side of (\ref{NewUP2}) starts with $1/\max_{n,k}\{|u_k(n)|^2\} \ge 1$ and ends with $N/\sum_{p=1}^N s^2(p) \le N$, having in mind that the sum of the  $N$ largest $u_k^2(n)$ values is at least equal to the eigenvector column (unity) energy.   

%on the left side of (\ref{NewUP2}) represents a line from $1$ to $N$ and

%This inequality results in the support uncertainty principle 
%\begin{gather}
%||\mathbf{x}||_0 ||\mathbf{X}||_0 \ge \frac{1}{(\displaystyle{\mathrm{MxAvg}}\{|u_k(n)|\})^2}, \label{NewUP}
%\end{gather}
%where 
%\begin{gather}\mathrm{MxAvg}\{|u_k(n)|\}=\displaystyle{\max_{\mathbb{M},\mathbb{K}}} \Big\{ \frac{1}{MK}\sum_{n \in \mathbb{M}}\sum_{k \in \mathbb{K}} |u_k(n)| \Big\} \label{MaxAv}
%\end{gather}
%denotes the maximal average value of $|u_k(n)|$ over the support for the signal and its GFT, $n \in \mathbb{M}$ and $k \in \mathbb{K}$. In general, this relation cannot be directly solved since the average on the right side is over the nonzero value indices of $x(n)$ and $X(k)$ that are not known. However, we can easily improve the support uncertainty principle bound through a computationally simple iterative procedure. 
%
%The trivial solution to (\ref{MaxAv}) is for $M=1$, $N=1$, when $||\mathbf{x}||_0 ||\mathbf{X}||_0=1$, and the value of (\ref{MaxAv}) is $\mathrm{MxAvg}\{|u_k(n)|\}=\displaystyle{\max_{n,k}}\{|u_k(n)|\}$, when (\ref{uncMM}) follows. However, if $1/\displaystyle{\max_{n,k}}\{|u_k(n)|\}>1$ then the inequality (\ref{uncMM}) does not allow $M=1$, $N=1$, and the smallest integer that satisfies the inequality in (\ref{uncMM}) should be used in (\ref{MaxAv}).
%

The computation complexity of this search can be reduced using an algorithm that will be presented next. 
The basic idea for the algorithm comes from the fact that the uncertainty bound  $Q=1/\max_{n,k}\{|u_k(n)|^2\}$ for the product $||\mathbf{x}||_0 ||\mathbf{X}||_0$ means that the smallest possible value of $||\mathbf{x}||_0 ||\mathbf{X}||_0$ is the  nearest, greater or equal, integer of $Q$, denoted by $\lceil Q \rceil$. This value is obtained as if all the terms $|u_k(n)|$ for $n \in \mathbb{M}$ and $k \in \mathbb{K}$ in (\ref{ImUB}) were equal to $\max_{n,k}\{|u_k(n)|^2\}$. However, the maximum possible value of the sum in (\ref{NewUP1}) is equal to the average value of the $\lceil Q \rceil$ largest $|u_k(n)|^2$, meaning that the bound is larger than $Q$ and should be corrected according the following  \textbf{Algorithm:}

\smallskip
 
\noindent \textbf{Step 0:}  Sort the absolute values of the transformation matrix elements, $|u_k(n)|$, into a nonincreasing order 
\begin{gather*}s(p)=\mathrm{sort}_{n,k}\{|u_k(n)|\},
\end{gather*}
with $n, k =0,1,\dots,N-1$, and $p=1,2,\dots, N^2$.

\smallskip

\noindent \textbf{Step 1:}  Calculate the bound in (\ref{uncMM}),  $Q=1/{\displaystyle{\max_{n,k}}}\{|u_k(n)|^2\}$, 
and its nearest, greater or equal, integer (ceiling of $Q$)
$$\lceil Q \rceil=\lceil \frac{1}{\max_{n,k}\{|u_k(n)|^2\}}\rceil,$$ 
being the minimum possible candidate for $||\mathbf{x}||_0 ||\mathbf{X}||_0$ value.

\smallskip

\noindent \textbf{Step 2:} Since the smallest possible integer for $||\mathbf{x}||_0 ||\mathbf{X}||_0$ is $\lceil Q \rceil$, recalculate the bound with $\lceil Q \rceil$  largest absolute values $|u_k(n)|$, instead of $\max_{n,k}\{|u_k(n)|^2\}$, according to (\ref{NewUP2}), 
\begin{gather} Q_N =  \frac{1}{\frac{1}{\lceil Q \rceil}\sum_{p=1}^{\lceil Q \rceil} s^2(p)}.\label{QN}
\end{gather}

\noindent \textbf{Step 3:} If $\lceil Q \rceil\ge  Q_N $
holds, then stop the algorithm, since inequality (\ref{NewUP2}) holds, and the uncertainty principle bound is 
\begin{gather}
||\mathbf{x}||_0 ||\mathbf{X}||_0 \ge  Q_N .
\end{gather}
If $\lceil Q \rceil < Q_N $, the inequality in (\ref{NewUP2}) does not hold. Set $ Q =  Q_N$ and go back to Step 2.

\noindent \textbf{Special case:} Consider the classical DFT analysis as a special case. The basis functions (eigenvectors with elements $u_k(n)$) are such that $|u_k(n)|=1/\sqrt{N}$. The average value  (\ref{NewUP1}) is constant for any set of $n,k$ and the standard DFT support uncertainty principle in (\ref{ClassUP}) follows. The presented algorithm is stopped in the first iteration since $Q=\lceil Q_N \rceil=N$.

\smallskip

\noindent \textbf{Comments on the algorithm.} Note that the average value in  (\ref{NewUP1}) is such that
\footnotesize
\begin{gather}
 \hspace{-1.5mm}||\mathbf{x}||_0 ||\mathbf{X}||_0 \!\ge\! \frac{1}{\displaystyle{\mathrm{Avg}}\{|u_k(n)|^2\}} \! \ge \! \frac{1}{\frac{1}{\lceil Q \rceil}\sum_{p=1}^{\lceil Q \rceil} s^2(p)} \! \ge \! \frac{1}{\max_{n,k}\{|u_k(n)|^2\}},\nonumber
\end{gather}
\normalsize
meaning that the bound in (\ref{QN}) satisfies (\ref{NewUP1}), but it could be tighter than (\ref{uncMM}). The proposed uncertainty principle bound in (\ref{NewUP1}) is always greater or equal to the bound in (\ref{uncMM}).

If we used the \textit{equality condition} in the Schwartz inequality, from (\ref{lwbound}) to (\ref{UnCElad0}),  which reads $\sqrt{|u_k(n)|}|x(n)|=c\sqrt{|u_k(n)|}|X(k)|$, for all $n,k$, the unit energy signal and its GFT should be constant $|x(n)|=1/\sqrt{M}$ and $|X(k)|=1/\sqrt{K}$, as in  \cite{elad2001generalized}. Then, relation (\ref{UnCElad0}) results in a tighter bound, 
\begin{gather}
||\mathbf{x}||_0 ||\mathbf{X}||_0 \!\ge\! \frac{1}{\Big(\displaystyle{\mathrm{Avg}}\{|u_k(n)|\}\Big)^2}\!\ge\! \frac{1}{\displaystyle{\mathrm{Avg}}\{|u_k(n)|^2\}}.
\end{gather}
In this case, we can use the same presented algorithm for the bound calculation, with
\begin{gather}
||\mathbf{x}||_0 ||\mathbf{X}||_0 
\ge \frac{1}{\Big(\frac{1}{||\mathbf{x}||_0 ||\mathbf{X}||_0}\sum_{p=1}^{||\mathbf{x}||_0 ||\mathbf{X}||_0} s(p)\Big)^2}
, \label{NewUP3}
\end{gather}
and $Q_N  = 1/\Big(\frac{1}{\lceil Q \rceil}\sum_{p=1}^{\lceil Q \rceil} s(p)\Big)^2$ in (\ref{QN}).

%\medskip
%
%
%(2) Assume now that there exist an eigenvector such that $u_{k_0}(n)=\delta(n-n_0)$. Then $\displaystyle{\max_{n,k}}\{|u_k(n)|^2\}=1$ meaning that $Q=1=\lceil Q \rceil$ and  
%$s^2(1)=\displaystyle{\max_{n,k}}\{|u_k(n)|^2\},$
%with $\lceil Q_N \rceil=\lceil Q \rceil=1$ in the Algorithm. This relation means that it is possible to get the relation $||\mathbf{x}||_0 ||\mathbf{X}||_0=1$, and that this bound can be achieved, producing the result in (\ref{uncMM}).

%(3) Finally, consider the case when  $1/\displaystyle{\max_{n,k}}\{|u_k(n)|^2\}=\lceil Q \rceil>1$ and $$s(1)=s(2)=\dots=s(\lceil Q \rceil)=\displaystyle{\max_{n,k}}\{|u_k(n)|\},$$ then, according to (\ref{QN}), $\lceil Q_N \rceil=\lceil Q \rceil$ and (\ref{uncMM}) holds.

\medskip

Using the well-known relation between the arithmetic and the geometric mean,   we can also write
\begin{gather}
||\mathbf{x}||_0 + ||\mathbf{X}||_0 \ge 2\sqrt{||\mathbf{x}||_0 ||\mathbf{X}||_0 } \ge \frac{2}{\frac{1}{\lceil Q \rceil}\sum_{p=1}^{\lceil Q \rceil} s(p)}. \nonumber
\end{gather}
This relation can be used to lower the reconstruction bounds in  compressive sensing \cite{donoho2006compressed,elad2001generalized,stankovic2020demystifying}.

 \section{Numerical Examples}\label{SecNE}

\begin{figure}
	[ptb]
	\begin{center}
		\includegraphics[scale=0.8]{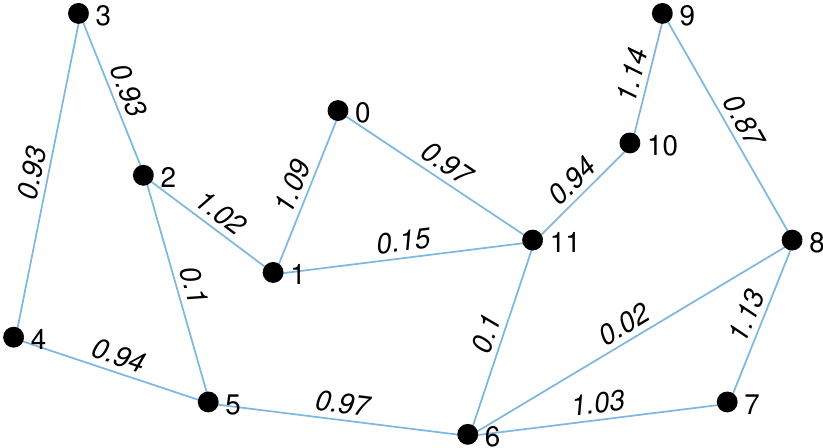}
		
		\vspace{3mm}
		
		\includegraphics[scale=0.90]{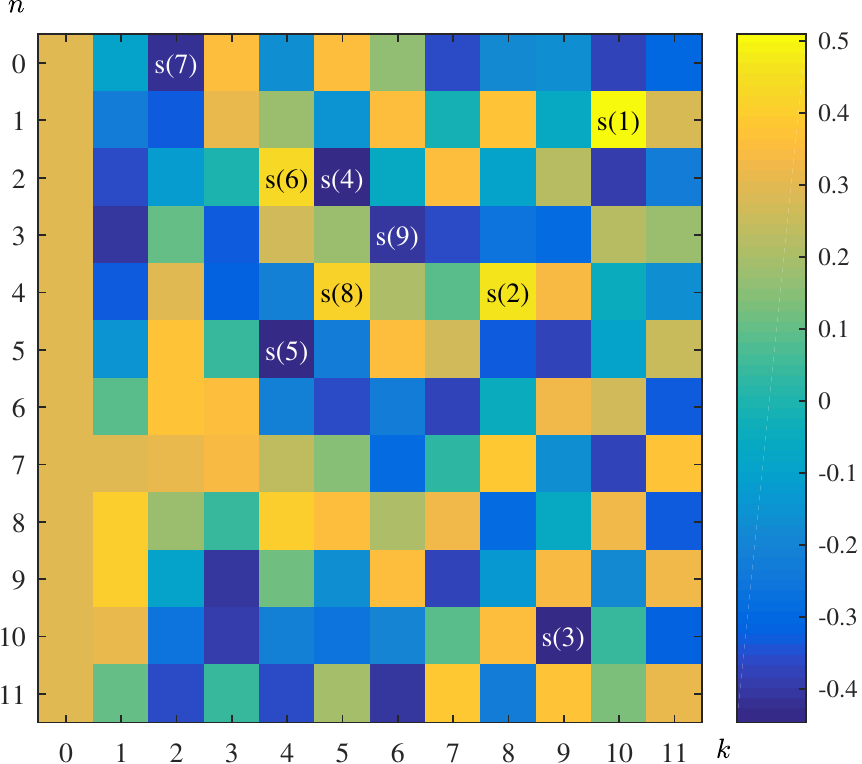}
		\caption{A graph (top) and its transformation matrix $\mathbf{U}$ with elements $u_k(n)$ (bottom). Several largest sorted values of $|u_k(n)|$, denoted by $s(p)$, are marked (the largest negative values are marked in white).  }\label{gggggg}\end{center}
\end{figure}

\noindent\textbf{Example 1.} Consider the graph with $N=12$ vertices, as in Fig. \ref{gggggg}(top). Its graph Laplacian is calculated, along with  the corresponding eigenvalues and eigenvectors, whose elements are $u_k(n)$, shown in Fig. \ref{gggggg}(bottom). 

%\begin{tabular}
%s(1) & s(2) & s(3) & s(4) & s(5) & s(5) & s(6) \\
%0.2597  &  0.2039  &  0.1991 &   0.1948  &  0.1878  &  0.1810 &   0.1778 
%\end{tabular}

The maximal value of the eigenvector elements is
$$\max_{n,k}\{|u_k(n)|^2\}=0.2597,$$
meaning that the uncertainty principle relation (\ref{uncMM}) yields
$$||\mathbf{x}||_0 ||\mathbf{X}||_0 \ge 3.8510=Q.$$
This relation states that the product of the numbers of nonzero elements in  $\mathbf{x}$ and $\mathbf{X}$ cannot be lower than $\lceil Q \rceil = 4$. The Step 2 in the algorithm, with $\lceil Q \rceil = 4$,  produces $\lceil Q_N \rceil = \lceil 4.6645  \rceil =5 $, meaning that the bound can be improved. In the next iteration, by letting  $\lceil Q \rceil = 5$, the value $\lceil Q_N \rceil = \lceil 4.7832 \rceil =5 $ is obtained, and the iteration process is stopped. The support uncertainty principle is now 
$$||\mathbf{x}||_0 ||\mathbf{X}||_0 \ge 4.7832.$$

The improvement in the support uncertainty principle bound is from $3.8510$ to $4.7832$. 

%Next, consider a graph with a small deviation from the circular graph, with one additional vertex $n=17$, Fig. \ref{gggggg_2}. Its transformation matrix $\mathbf{U}$ is shown in this figure as well. The uncertainty principle (\ref{uncMM}) for this graph is extremely low, 
%$$||\mathbf{x}||_0 ||\mathbf{X}||_0 \ge 1.2703,$$
%just above the trivial bound equal to $1$.
%
%Using (\ref{NewUP}) and the presented iterative procedure 
%we get
%$$||\mathbf{x}||_0 ||\mathbf{X}||_0 \ge 4.5081.$$
%
%The bound improvement here is large. Note that the support uncertainty bound for the unweighted and undirected circular graph with $N=16$ would be $8$. 
%
%\begin{figure}
%	[ptb]
%	\begin{center}
%		\includegraphics[scale=0.8]{gggggg_2.eps}
%		 
%		 \vspace{5mm}
%		 
%			\includegraphics[scale=0.95]{gggggg_2U.eps}
%		\caption{A graph and its transformation matrix $\mathbf{U}$ with elements $u_k(n)$. }\label{gggggg_2}\end{center}
%\end{figure}  

%
%The support form of the uncertainty principle for the Rihaczek distribution  follows  from 
%\begin{gather}\sum_{n \in \mathbb{M}}\sum_{k \in \mathbb{K}}|x(n) X(k) u_k(n)|^0 = ||\mathbf{E}||_0 \le \sum_{n \in \mathbb{M}}\sum_{k \in \mathbb{K}}|x(n)|^0 |X(k)|^0 =  ||\mathbf{x}||_0 ||\mathbf{X}||_0.\end{gather}

\medskip
%
%\noindent \textbf{Windowed and filtered forms.} If the GFT is filtered by a band-pass filter defined by the diagonal matrix $H(\boldsymbol{\Lambda})$ whose width (nonzero diagonal elements) covers $P$ GFT values. This means that $$||H(\boldsymbol{\Lambda})\mathbf{X}||_0 \le P.$$
%
%The local GFT is then obtained as the inverse GFT of the product $H(\boldsymbol{\Lambda})\mathbf{X}$, that is 
%\begin{gather}
%\mathbf{s}=\mathbf{U} (H(\boldsymbol{\Lambda})\mathbf{X}).
%\end{gather}
%The support uncertainty principle for the LGFT, $\mathbf{s}$, is 
%\begin{gather}
%||\mathbf{s}||_0  \ge \frac{1}{||H(\boldsymbol{\Lambda})\mathbf{X}||_0 \ \ \displaystyle{\max_{n,k}}\{|u_k(n)|^2\}} \ge \frac{1}{P \ \ \displaystyle{\max_{n,k}}\{|u_k(n)|^2\}} 
%\end{gather}
%
%The same form holds for the vertex domain windowed graph signals
%\begin{gather}
%||\mathbf{s}||_0  \ge \frac{1}{||\mathbf{Wx}||_0 \ \ \displaystyle{\max_{n,k}}\{|u_k(n)|^2\}} \ge \frac{1}{W \ \ \displaystyle{\max_{n,k}}\{|u_k(n)|^2\}} 
%\end{gather}
% 
% \noindent \textbf{Equality Conditions for the Windowed and filtered forms.}

\begin{figure}
	[ptb]
	\begin{center}
		\includegraphics[scale=0.8]{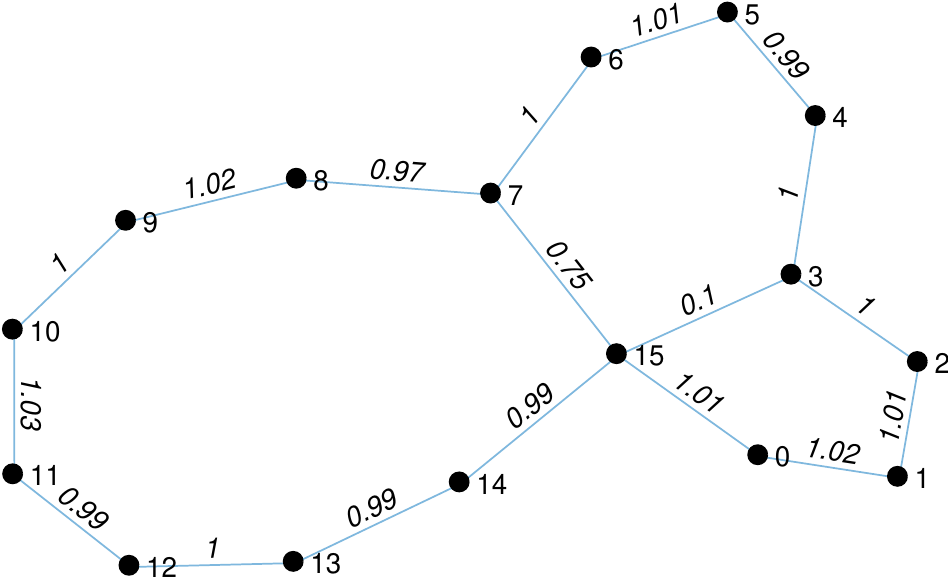}
		
		\vspace{3mm}
		
		\includegraphics[scale=0.9]{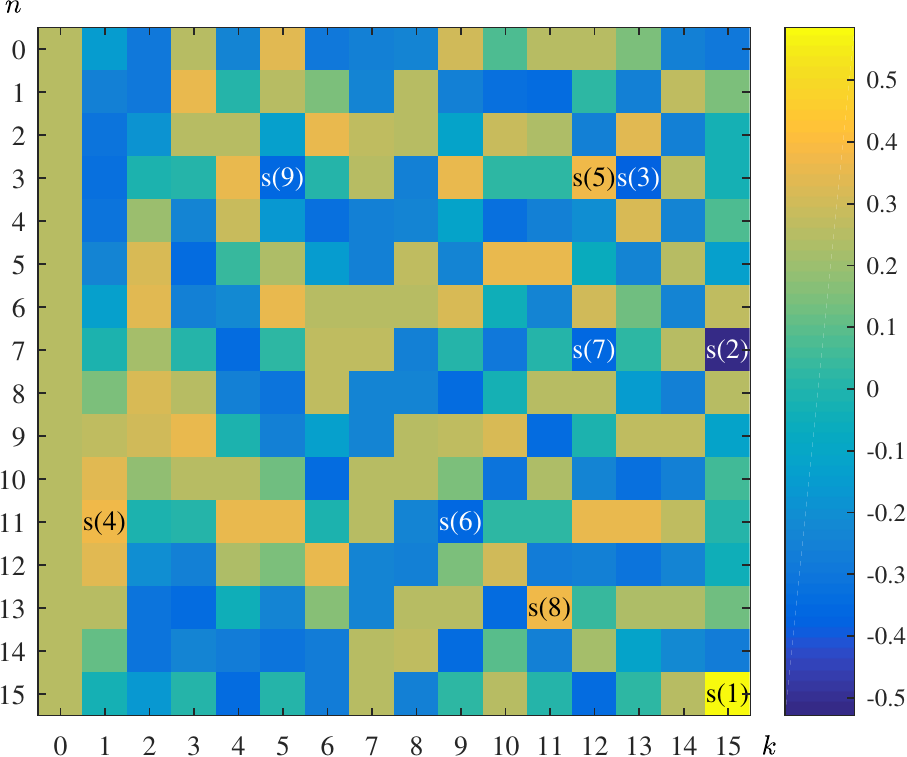}
		\caption{A graph (top) and its transformation matrix $\mathbf{U}$ with elements $u_k(n)$ (bottom). Several largest sorted values of $|u_k(n)|$, denoted by $s(p)$, are marked (the largest negative values are marked in white). }\label{gggggg_3}\end{center}
\end{figure}  

\noindent\textbf{Example 2.} Consider the graph with $N=16$, shown in Fig. \ref{gggggg_3}(top). Its transformation matrix $\mathbf{U}$ is given in Fig. \ref{gggggg_3}(bottom). The uncertainty principle (\ref{uncMM}) for this graph is quite low, 
$$||\mathbf{x}||_0 ||\mathbf{X}||_0 \ge 2.9156,$$
slightly above the trivial bound equal to $1$.
Several the largest absolute values, $|u_k(n)|$, of the transformation matrix $\mathbf{U}$ are
$$\mathbf{s}=[0.5857, \ 0.5285, \   0.3743,  \  0.3669, \   0.3659, \ 0.3658, \  \dots ],$$
as shown in  Fig. \ref{gggggg_3}(bottom). 
The largest  squared absolute value is $\max_{n,k}\{|u_k(n)|^2\}=0.3430$, producing 
$$\lceil Q \rceil=\lceil \frac{1}{\displaystyle{\max_{n,k}}\{|u_k(n)|^2\}}\rceil=\lceil 2.9151 \rceil=3.$$ 
Now the iterative procedure is started from Step 2 in the algorithm, with $\lceil Q \rceil=3$.  The presented iterative procedure produced $\lceil Q_N \rceil= \lceil 4.4590 \rceil=5$ in the first iteration, then $\lceil Q_N \rceil= \lceil 4.8499 \rceil=5$ in the second iteration, when the iteration process is stopped since the value of $\lceil Q \rceil$ was not changed. The final result of this iterative procedure is the improved support uncertainty principle bound,
$$||\mathbf{x}||_0 ||\mathbf{X}||_0 \ge 4.8499.$$

 \smallskip
 
\noindent\textbf{Example 3.} An unweighted, undirected, large circular graph with $N=5000$ vertices, is modified in such a way that the vertices $n=2499$ and $m=4999$ are connected with a unit weight, $W_{2499,4999}=W_{4999,2499}=1$. For this graph, the bound in (\ref{uncMM}) is $1/\max_{n,k}\{|u_k(n)|^2\}=2.8$, just slightly greater than $1$, while the proposed method in (\ref{NewUP2})  produces $||\mathbf{x}||_0 ||\mathbf{X}||_0 \ge  17.99$ (or $||\mathbf{x}||_0 ||\mathbf{X}||_0 \ge  2244.3$ with (\ref{NewUP3})). By reducing the added edge weight value  to $0.1$, then to $0.01$, and finally to $0.0001$, the respective bounds obtained with (\ref{NewUP2}), $116.97$, $689.99$, and $2483$, approach to the pure \textit{undirected} circular graph bound, equal to $N/2=2500$.

\section{Conclusion}The uncertainty principle of graph signals is revisited using the graph Rihaczek distribution as an analysis tool. This derivation is used as the basis to introduce improved bounds for the uncertainty principle.  The improved bounds can be used in compressive sensing to lower the coherence index-based reconstruction sparsity bound.

\medskip

\noindent\textbf{Acknowledgments.} The author is thankful to Prof. Milo\v{s} Dakovi\'{c} and Dr. Milo\v{s} Brajovi\'{c} for constructive comments.

%\section*{References}

%\nocite{*}
\bibliographystyle{ieeetr}

\bibliography{graph-signal-processing}

\begin{thebibliography}{10}

\bibitem{boashash2015time}
B.~Boashash, {\em Time-frequency signal analysis and processing: {A}
  comprehensive reference}.
\newblock Academic Press, 2015.

\bibitem{cohen1995time}
L.~Cohen, {\em Time-frequency Analysis}.
\newblock Prentice Hall PTR, 1995.

\bibitem{stankovic1997highly}
L.~Stankovic, ``Highly concentrated time-frequency distributions: Pseudo
  quantum signal representation,'' {\em IEEE Transactions on Signal
  Processing}, vol.~45, no.~3, pp.~543--551, 1997.

\bibitem{donoho2006compressed}
D.~L. Donoho, ``Compressed sensing,'' {\em IEEE Transactions on information
  theory}, vol.~52, no.~4, pp.~1289--1306, 2006.

\bibitem{ricaud2014survey}
B.~Ricaud and B.~Torr{\'e}sani, ``A survey of uncertainty principles and some
  signal processing applications,'' {\em Advances in Computational
  Mathematics}, vol.~40, no.~3, pp.~629--650, 2014.

\bibitem{stankovic2001measure}
L.~Stankovi{\'c}, ``A measure of some time--frequency distributions
  concentration,'' {\em Signal Processing}, vol.~81, no.~3, pp.~621--631, 2001.

\bibitem{perraudin2018global}
N.~Perraudin, B.~Ricaud, D.~I. Shuman, and P.~Vandergheynst, ``Global and local
  uncertainty principles for signals on graphs,'' {\em APSIPA Transactions on
  Signal and Information Processing}, vol.~7, no.~e3, pp.~1--26, 2018.

\bibitem{elad2001generalized}
M.~Elad and A.~M. Bruckstein, ``Generalized uncertainty principle and sparse
  representation in pairs of bases,'' {\em IEEE Transactions on Information
  Theory}, vol.~48, no.~9, pp.~2558--2567, 2002.

\bibitem{pasdeloup2019uncertainty}
B.~Pasdeloup, V.~Gripon, R.~Alami, and M.~G. Rabbat, ``Uncertainty principle on
  graphs,'' in {\em Vertex-Frequency Analysis of Graph Signals}, pp.~317--340,
  Springer, 2019.

\bibitem{Tsitsvero2016}
M.~Tsitsvero, S.~Barbarossa, and P.~{Di Lorenzo}, ``Signals on graphs:
  Uncertainty principle and sampling,'' {\em IEEE Transactions on Signal
  Processing}, vol.~64, no.~18, pp.~539--554, 2016.

\bibitem{Agaskar}
A.~Agaskar and Y.~M. Lu, ``A spectral graph uncertainty principle,'' {\em IEEE
  Transactions on Information Theory}, vol.~59, no.~7, pp.~4338--4356, 2013.

\bibitem{stankovic2019vertexTEL}
L.~Stankovi{\'c} and E.~Sejdi{\'c}, {\em Vertex-Frequency Analysis of Graph
  Signals}.
\newblock Springer, 2019.

\bibitem{stankovic2019graph}
L.~Stankovic, D.~Mandic, M.~Dakovic, M.~Brajovic, B.~Scalzo, and
  T.~Constantinides, ``Graph signal processing--{P}art {I}: Graphs, graph
  spectra, and spectral clustering,'' {\em arXiv preprint arXiv:1907.03467},
  2019.

\bibitem{stankovic2018vertex}
L.~Stankovi{\'c}, E.~Sejdi{\'c}, and M.~Dakovi{\'c}, ``Vertex-frequency energy
  distributions,'' {\em IEEE Signal Processing Letters}, vol.~25, no.~3,
  pp.~358--362, 2018.

\bibitem{stankovic2020demystifying}
L.~Stankovic, D.~P. Mandic, M.~Dakovic, and I.~Kisil, ``Demystifying the
  coherence index in compressive sensing [lecture notes],'' {\em IEEE Signal
  Processing Magazine}, vol.~37, no.~1, pp.~152--162, 2020.

\end{thebibliography}

\end{document}